\documentclass[aps,prd,twocolumn,noshowpacs,nofootinbib,superscriptaddress,groupedaddress,
floatfix,superscriptaddress, preprintnumbers]{revtex4-1}%
\usepackage{listings}
\usepackage[utf8]{inputenc}
\usepackage{amsmath}
\usepackage{amsfonts}
\usepackage{graphicx}
\usepackage{amssymb}
\usepackage{array,multirow}%
\usepackage{color}
\begin{document}

\title{Heavy Quarkonia, Heavy-Light Tetraquarks and the Chiral Quark-Soliton Model}
\author{Maciej Kucab}
\email{maciej.kucab@student.uj.edu.pl}
\affiliation{Institute of Theoretical Physics, Jagiellonian University, S. \L ojasiewicza 11, 30-348 Krak\'ow, Poland}
\author{Micha\l~Prasza\l owicz}
\email{michal.praszalowicz@uj.edu.pl}
\affiliation{Institute of Theoretical Physics, Jagiellonian University, S. \L ojasiewicza 11, 30-348 Krak\'ow, Poland}

\begin{abstract}
We apply the Chiral Quark-Soliton Model used previously to describe baryons with one heavy quark to the case of heavy tetraquarks.
We argue that the model is insensitive to the nature of the heavy object bound by the soliton, {\em i.e.} to its mass and spin. Therefore, a heavy quark 
can be replaced by an anti-diquark without modifying the soliton background. Diquark dynamics is taken into account by means of
the non-relativistic Schrödinger equation with the Cornell potential. We fix the Cornell potential parameters from the charmonia and bottomia
spectra. We first compute $B_c$ meson masses to check our fitting procedure, and then compute diquark masses by appropriately 
rescaling color factors in the Cornell potential. We then compute tetraquark masses and confirm previous findings that only $bb$ tetraquarks are bound.

\end{abstract}
\maketitle

\section{Introduction}

In 2022 the LHCb Collaboration discovered the doubly charmed tetraquark ${\cal{T}}_{cc}$
\cite{LHCb:2021vvq,LHCb:2021auc}  in
$D^0D^0\pi^+$ invariant mass distribution. ${\cal{T}}_{cc}$ mass of 3875~MeV is just below
the $D^0D^*$ threshold. The LHCb discovery triggered theoretical activity. We refer the reader to
a review on
multiquark states,
both experimental and theoretical, before ${\cal T}_{cc}^+$ discovery \cite{Karliner:2017qhf}
and after the LHCb paper  \cite{Chen:2022asf} (and references therein).

Motivated by ${\cal T}_{cc}^+$  discovery one of us proposed a model \cite{Praszalowicz:2022sqx} 
where
heavy tetraquarks $\mathcal{T}_{QQ}$ were  described as a chiral soliton
 and a $\bar{Q}\bar{Q}$ diquark. The Chiral
Quark Soliton Model ($\chi$QSM) has been formulated to describe light baryons
(see \cite{Diakonov:1987ty} and
Refs.~\cite{Christov:1995vm,Alkofer:1994ph,Petrov:2016vvl} for review) where
the soliton is constructed from $N_{c}$ light quarks. It has been argued in
Refs.~\cite{Yang:2016qdz,Kim:2017jpx,Kim:2017khv,Polyakov:2022eub,Praszalowicz:2022hcp}
that in the large $N_{c}$ limit mean chiral fields of the soliton do not
change if one valence quark is replaced by a heavy quark $Q$. Such a
replacement leads to a successful phenomenological description of baryons with
one heavy quark~\cite{Yang:2016qdz,Kim:2017khv,Polyakov:2022eub}. Since after removing one light quark the soliton is in a
color $\overline{\mathbf{3}}$ (or more precisely a color representation
$\mathcal{R}$ corresponding to an antisymmetric product on $N_{c}-1$ quarks),
adding a heavy quark in color \textbf{3} leads to multiplets of heavy baryons
that are conveniently characterized by SU(3)$_{\mathrm{flavor}}$ quantum
numbers of $N_{c}-1$ light quarks (\emph{i.e.} a diquark for $N_{c}=3$). In
this respect the $\chi$QSM is identical to a quark model.

It has been shown in
Refs.~\cite{Yang:2016qdz,Kim:2017jpx,Kim:2017khv,Polyakov:2022eub,Praszalowicz:2022hcp}
that for a successful phenomenological description of heavy baryons, it is
enough to add the masses of a soliton and a heavy quark, and include a
spin-spin interaction between the two. The model describes well both charm and
bottom baryon spectra~\cite{Yang:2016qdz,Kim:2017khv,Polyakov:2022eub},
indicating that binding effects of the soliton-$Q$ system  do not
depend on the heavy quark mass. We present quantitative evidence
for this independence in Sect.~\ref{sec:chiQSM}.
This observation suggests that equally good
description should hold for a system where a heavy quark is replaced by a heavy
(anti)diquark $\bar{Q}_{1}\bar{Q}_{2}$ in color triplet. In
Ref.~\cite{Praszalowicz:2022sqx} and earlier in
Ref.~\cite{Praszalowicz:2019lje} one considered the case where $Q_{1}=Q_{2}$.

In the present paper we study a more general case where heavy
quarks\footnote{In what follows we will use term \emph{quark} or
\emph{diquark} referring both to $Q_{1},Q_{2}$ or $\bar{Q}_{1},\bar{Q}_{2}$.}
can be both identical or different, \emph{i.e.} we consider $cc$, $bb$ and $cb$
diquarks. Diquark dynamics is modeled by a non-relativistic Schr{\"o}dinger
equation with the Cornell potential~\cite{Eichten:1978tg,Mateu:2018zym} and
spin-spin interaction of heavy quarks, which has not been explicitly included
in Ref.~\cite{Praszalowicz:2022sqx}. Since we are interested only in the
diquark ground states, angular momentum and tensor terms are neglected. We
use as an input $J/\psi$, $\eta_{c}$,$\Upsilon$ and $\eta_{b}$ mesons to
constrain the Cornell potential parameters and quark masses. As a result masses of
$b\bar{c}$ or $\bar{c}b$ mesons are predictions and actually test our
approach. The model reproduces very well two known $B_{c}^{+}(1S_0,6274.5)$ and
$B_{c}^{\pm}(2S_0,6871.2)$ mesons~\cite{Workman:2022ynf}.

Once the Cornell potential parameters are fixed, we can compute the diquark masses
by coupling quark color charges to an anti-triplet rather than to a singlet,
as in the meson case. Finally, by adding the diquark mass to the soliton mass
with diquark-soliton spin interaction we obtain predictions for the tetraquark
masses. 

Two heavy quarks of the same flavor (say $cc$ or $bb$), can
form a color anti-triplet (antisymmetric in color) provided they are symmetric
in spin~\cite{Gelman:2002wf}. Therefore they form a tight object of spin 1.
Hence, two heavy antiquarks are  in color $\mathbf{3}$ and spin 1, behaving as
a spin 1 heavy \emph{quark}. Additionally, a $cb$ diquark can be in a state of spin zero,
which is antisymmetric in flavor.

Heavy tetraquarks have been anti{\-}cipated theoretically already many yers
ago~\cite{Carlson:1987hh,Manohar:1992nd} on the basis of heavy quark
symmetry~\cite{Isgur:1991wq}  (see also 
\cite{Navarra:2007yw,Cohen:2006jg,Esposito:2013fma,Cai:2019orb,Karliner:2017qjm,Agaev:2018vag,Agaev:2018khe}). 
Probably the first estimate of the tetraquark mass was done by Lipkin in 1986
\cite{Lipkin:1986dw}
(although the fourfold heavy tetraquarks were discussed even earlier in 1982 \cite{Ader:1981db}). 
A phenomenological analysis of heavy tetraquarks has been recently carried out
in Ref.~\cite{Eichten:2017ffp}. In fact our model is very reminiscent to the
one of Ref.~\cite{Eichten:2017ffp} where tetraquark mass formulas are
identical to those for heavy baryons, with some modification due to the
integer or zero spin of the heavy diquark.

Our findings can be summarized as follows. Diquark dynamics restricted to the $s$
channel,
modeled by the Cornell potential, describes well charmonia and bottomia ground
states and first excited states, however the value of the string tension giving the
best fit is different in $c$ and $b$ channels. This is consistent with global fits 
\cite{Mateu:2018zym}. Using the parameters fixed from meson spectra we compute
diquark masses and the tetraquark masses. We find that only $bb$ tetraquarks
are bound. 

In Sect.~\ref{sec:chiQSM} we introduce the $\chi$QSM and discuss its application to heavy baryons.
We present arguments that the soliton properties do not depend on the heavy quark mass.
Next, we introduce classification of the tetrquark states according to the SU(3) content of
the light subsystem, and derive pertinent mass formulas. In Sect.~\ref{sec:diquark} we
solve Schrödinger equation for heavy mesons and fix the Cornell potential parameters.
As a test we compute $B_c$ mesons masses, and then  the diquark masses.
Numerical results for the tetraquark masses are presented in Sect~\ref{sec:masses}.
We summarize our findings in Sect.~\ref{sec:summary}.

\section{Chiral Quark Soliton Model}

\label{sec:chiQSM}

In this section we briefly recall the main features of the $\chi$QSM,
\cite{Diakonov:1987ty,Christov:1995vm,Alkofer:1994ph,Petrov:2016vvl} (and
references therein). We first discuss application of the $\chi$QSM to heavy
baryons and then to tetraquarks.

\subsection{Heavy baryons}

\label{sec:HB}

The soliton in the current approach corresponds to a stable aggregate configuration 
of valence quarks and a fully occupied Dirac sea.
In the large $N_{c}$ limit, $N_{c}$ (or $N_{c}-1$) relativistic valence quarks
polarize the Dirac sea, which  in turn modifies the valence quark levels,
which in turn distort the sea, until a stable soliton configuration is
reached~\cite{Witten:1979kh,WittenCA}. Quantum numbers are generated by
quantization of zero modes, corresponding to the rotations in the SU(3) space and
in the configuration space.
In the chiral limit the soliton energy is given by a formula analogous to the
quantum mechanical symmetric top~\cite{Guadagnini:1983uv,Mazur:1984yf,Jain:1984gp}%
\begin{equation}
E_{\text{sol}}=M_{\mathrm{sol}}+\frac{J(J+1)}{2I_{1}}+\frac{C_{2}%
(p,q)-J(J+1)-3/4\,Y^{\prime\,2}}{2I_{2}}.\label{eq:Erot}%
\end{equation}
Here $M_{\mathrm{sol}}$ is a classical soliton mass, $I_{1,2}$ stand for the
moments of inertia, $C_{2}(p,q)$ is the SU(3) Casimir for the baryon multiplet
and $J$ corresponds to the soliton spin. In the case of $N_{c}-1$ valence
quarks  $Y^{\prime}=(N_{c}-1)/3=2/3$ in a real world, and the allowed SU(3)
representations are $\boldsymbol{\overline{3}}$ with spin $J=0$ and
$\boldsymbol{6}$ with spin $J=1$~\cite{Yang:2016qdz}.

Hamiltonian (\ref{eq:Erot}) has to be supplemented by the chiral symmetry
breaking part, which can be found in Ref.~\cite{Blotz:1992pw} and by  the hyperfine splitting
part~\cite{Yang:2016qdz}
\begin{equation}
H_{SQ}=\frac{2}{3}\frac{\varkappa}{m_{Q}} {\boldsymbol{J}}\cdot{\boldsymbol{S}%
}_{Q} \label{eq:ssinter}%
\end{equation}
where ${\boldsymbol{J}}$ and ${\boldsymbol{S}}_{Q}$ stand for the soliton and
the heavy quark or diquark spin, respectively. Since the spin of the
$\boldsymbol{\overline{3}}$ representation is zero, there is no hyperfine
splitting in this case. Chiral symmetry breaking part leads to the mass
splittings proportional to the baryon hypercharge, denoted below by
$\delta_{\boldsymbol{\overline{3}},\boldsymbol{6}}$~\cite{Yang:2016qdz}.

Mass formulas for heavy baryons read therefore as
follows~\cite{Yang:2016qdz,Praszalowicz:2019lje}:
\begin{align}
M_{B_Q,\overline{\boldsymbol{3}}}=m_{Q}+M_{\mathrm{sol}} & +\frac{1}{2I_{2}%
}+\delta_{\boldsymbol{\overline{3}}}Y_{B}\, , \nonumber\\
M_{B_Q,\boldsymbol{6},s} =m_{Q}+M_{\mathrm{sol}} & +\frac{1}{2I_{2}} +\frac
{1}{I_{1}}+\delta_{\boldsymbol{6}}Y_{B}\nonumber\\
&  +\frac{\kappa}{m_{Q}}\left\{
\begin{array}
[c]{ccc}%
-2/3 & \text{for} & s=1/2\\
\, & \, & \,\\
+1/3 & \text{for} & s=3/2
\end{array}
\right.  \, . \label{eq:M3barM6mass}%
\end{align}
Here $Y_{B}$ stands for a hypercharge of a given baryon. In the case of
anti-triplet soliton spin $J=0$ and the corresponding heavy baryons have spin
1/2, for sextet $J=1$ and the corresponding baryons have spin 1/2 and 3/2.

It has been shown in  Refs.~\cite{Yang:2016qdz,Kim:2017khv,Polyakov:2022eub}  that the above mass formulas lead to
a very good description of heavy baryon spectra. Below we examine the main
features of our approach:
\begin{enumerate}
  \setlength\itemsep{0.005em}
 \item soliton properties are independent of the heavy quark mass,
\item soliton properties do not depend on the spin coupling between a soliton and a heavy quark,
\item hyperfine splittings are proportional to $1/m_{Q}$.
\end{enumerate}

Averaging over spin and hypercharge we define mean anti-triplet and
sextet masses: 
\begin{eqnarray}
M_{\mathbf{\overline{3}}}^{Q}& = &m_{Q}+M_{\mathrm{sol}}  +\frac{1}{2I_{2}} =  \left. 2408.2\right| _{c} = \left.5736.2\right| _{b} \, ,
 \\
M_{\mathbf{6}}^{Q} & = & m_{Q}+M_{\mathrm{sol}}  +\frac{1}{2I_{2}} +\frac{1}{I_{1}}= \left. 2579.4 \right| _{c} 
=  \left. 5906.5 \right| _{b} \notag
\label{eq:repave}
\end{eqnarray}
in MeV.

As it was discussed in Ref.~\cite{Praszalowicz:2022sqx} one can form differences of average multiplet
masses between the $b$ and $c$ sectors to compute heavy quark mass difference
(in MeV):
\begin{equation}
m_{b}-m_{c}=\left.  3328\right| _{\overline{\boldsymbol{3}}}=\left.
3327\right| _{{\boldsymbol{6}}}\, ,\label{eq:mQdiff}%
\end{equation}
which illustrates properties 1 and 2 above.

Furthermore, one can estimate the hyperfine splitting parameter enetering
(\ref{eq:ssinter}):
\begin{align}
\frac{\varkappa}{m_{c}}  & = \left.  64.6\right| _{\Sigma_{c}} = \left.
67.2\right| _{\Xi_{c}} = \left.  70.7\right| _{\Omega_{c}} \, , \cr \frac
{\varkappa}{m_{b}} & = \left.  19.4 \right| _{\Sigma_{b}} = \left.
18.8\right| _{\Xi_{b}}\label{eq:spintest}%
\end{align}
(in MeV). From these estimates we get
\begin{equation}
\frac{m_{c}}{m_{b}}\simeq0.27 \div0.30\label{eq:mQratio}%
\end{equation}
with the average value of $0.283$, which is close to the PDG value of
0.3~\cite{Workman:2022ynf} in agreement with properties 2 and 3.

From Eqs.~(\ref{eq:mQdiff}) and (\ref{eq:mQratio}) one can estimate heavy
quark masses
\begin{align}
m_{c}  &  =1206\div1426\;\text{MeV,}\nonumber\\
m_{b}  &  =4533\div4753\;\text{MeV} \, , \label{eq:mQbaryon}%
\end{align}
which are a bit higher (especially $m_{b}$) than in the PDG~\cite{Workman:2022ynf}%
. For $m_{c}/m_{b}=0.283$ we get $m_{c}=1314.1$~MeV and $m_{b}=4641.5$~MeV,
which is still lower than the effective values used in
Ref.~\cite{Karliner:2014gca}. One should, however, remember that the quark
masses in effective models may differ from the QCD estimates in the
$\overline{\mathrm{MS}}$ scheme.

In Ref.~\cite{Yang:2016qdz} heavy quark dependence of the mass formulas 
(\ref{eq:M3barM6mass}) was tested by computing the non-strange moment of inertia from the
$\mathbf{6} - {\overline{\boldsymbol{3}}}$ average mass differences where both spin and hypercharge splittings
cancel:
\begin{align}
\frac{1}{I_{1}}=M_{\mathbf{6}}^{Q} - M_{\mathbf{\overline{3}}}^{Q} =\left.
171.2\right| _{c}=\left.  170.3\right| _{b}\label{eq:I1Q}%
\end{align}
in MeV. As we see from (\ref{eq:I1Q}) heavy quark masses cancel almost exactly,
which again illustrates properties 1 and 2.
We can therefore
safely assume that formulas (\ref{eq:mQdiff}) are valid for any heavy object in color triplet
replacing $Q$.

\subsection{Heavy Tetraquarks}

Since heavy tetraquarks in the $\chi$QSM are formed by replacing a heavy quark
by a diquark, and since the mass of the soliton is independent of the heavy
quark or diquark mass and spin, very simple tetraquark mass formulas emerge,
which relate tetraquark masses to the baryon masses \cite{Praszalowicz:2022sqx,Eichten:1978tg}. 
For the ground state
anti-triplet the mass formula is particularly simple, since the soliton in this case is
spinless and the hyper-fine splitting (\ref{eq:ssinter}) is not present
\begin{equation}
M_{\bar{Q}\bar{Q}}^{\text{tetra}\,\boldsymbol{\overline{3}}} =M_{B_{Q},\overline
{\boldsymbol{3}}} -m_{Q} +m_{\bar{Q}\bar{Q}} \, .
\label{eq:tetra3bar}
\end{equation}
Here $M_{B_{Q},\overline{\boldsymbol{3}}}$ stands for 
$\Lambda_{Q}( \left. 2286.5\right| _{c},  \left. 5619.6\right| _{b})$ or
isospin averaged
$\Xi_{Q}( \left. 2469\right| _{c},  \left. 5794.5\right| _{b})$ 
mass, $m_{\bar{Q}\bar{Q}}$ denotes the anti-diquark mass to
be discussed in Sect.~\ref{ssec:numQQ}, and $m_{Q}$ stands for the heavy quark mass.

In the case of sextet, since the soliton spin is $J=1$, we have to distinguish
two cases when the diquark spin is zero or one. It is convenient to introduce spin and isospin
averaged baryon masses:
\begin{eqnarray}
M_{B_Q,\boldsymbol{6}} &=&\frac{1}{2T+1}\sum_{T_3}\frac{1}{3}\left(  M_{B_Q,\boldsymbol{6},T_3,1/2} +2\,
M_{B_Q,\boldsymbol{6},T_3, 3/2} \right) \label{eq:MB6ave} \notag \\
 ~&~&
\end{eqnarray}
where $B_Q$ stands for $\Sigma_{Q}( \left. 2496.6\right| _{c},  \left. 5826\right| _{b})$, 
$\Xi'_{Q}( \left. 2623.2\right| _{c},  \left. 5947.6\right| _{b})$ or 
$\Omega_Q( \left. 2742.3\right| _{c},  \left.6065\right| _{b})$\footnote{For $\Omega_b^{\ast}$ we take mass estimate from Ref.~\cite{Yang:2016qdz}.} 
in MeV.
The mass formulas read as follows:
\begin{equation}
M_{\bar{Q}\bar{Q}}^{\text{tetra}\,\boldsymbol{6}} =M_{B_{Q},\boldsymbol{6}} -m_{Q} +
m_{\bar{Q}\bar{Q}} +C_{\rm spin}\frac{2}{3}\frac{\varkappa}{m_{Q}}\frac{m_{Q}%
}{m_{\bar{Q}\bar{Q}}}%
\label{eq:tetra6}
\end{equation}
where
\begin{equation}
C_{\rm spin}=\left\{
\begin{array}
[c]{ccc}%
\left.
\begin{array}
[c]{rcc}%
-2 & \text{for} & s=0\\
-1 & \text{for} & s=1\\
1 & \text{for} & s=2
\end{array}
\right\}   & \text{for} & S_{\bar{Q}\bar{Q}}=1\\%
\begin{array}
[c]{rcc}%
0 & \text{for} & s=0
\end{array}
& \text{for} & S_{\bar{Q}\bar{Q}}=0
\end{array}
\right.
\end{equation}

Mass formulas (\ref{eq:tetra3bar}) and (\ref{eq:tetra6}) relate tetraquark masses directly to heavy
baryon masses, and therefore are fairly model independent. They are analogous
to  Eq.(1) of Ref.~\cite{Eichten:2017ffp}. The spin part
has been discussed in \cite{Karliner:2021wju}, however, the hyper-fine
coupling has not been specified. Here we know the value of $\varkappa/m_{c,b}$
(\ref{eq:spintest}), so in order to estimate tetraquark masses we only need
heavy diquark mass $m_{\bar{Q}\bar{Q}}$ for $m_{Q}$ in the range
(\ref{eq:mQbaryon}).

Before proceeding to numerical calculations we need to know the strong decay thresholds
that depend on the $J^P$ quantum numbers, which are listed in Table~\ref{tab:thresholds}.

\section{Heavy Mesons and Diquarks}

\label{sec:diquark}

\subsection{Mass Formulas}

In order to predict heavy tetraquark masses one needs
a reliable estimate of the heavy diquark mass. Following
Ref.~\cite{Praszalowicz:2022sqx}
we use a non-relativistic Schr{\"o}dinger equation with the
Cornell potential~\cite{Eichten:1978tg,Mateu:2018zym}
\begin{equation}
V(r)=-\frac{\kappa}{r}+\sigma\,r + \frac{2}{3}C_{\rm color}\frac{\alpha _{\text{s}}}{m_{1}m_{2}r^{2}}%
(s_{1}\cdot s_{2})\,\delta (r),
\label{eq:Cornell}%
\end{equation}
including spin-spin interaction, which we treat as a perturbation. 
Since we are interested in $s$ wave states only,
we do not include tensor and spin-orbit interactions. Here $m_{1,2}$ stand
for heavy quark masses and we also introduce a reduced mass
\begin{equation}
\mu =\frac{m_{1}m_{2}}{m_{1}+m_{2}} \, ,
\label{eq:redmass}
\end{equation}%
which is equal to $m/2$ for quarks of identical mass $m$. String tension $\sigma$
should be in principle a universal constant, however it is known from global analyses
that good quality fits require $\sigma$, which is different in the $c$ and $b$ sector \cite{Mateu:2018zym}.
Since the Coulomb part  follows from the one gluon exchange $\kappa=C_{\rm color} \alpha_{\mathrm{s}}$,
where $C_{\rm color}$ is a color factor.

Here we adopt 
units where dim[$m$]=GeV, dim[$r$]=1/GeV, dim[$\sigma$]=GeV$^2$ and $\kappa$
is dimensionless.

\renewcommand{\arraystretch}{1.3} 
\begin{widetext} 
\begin{center}
\begin{table}[tbp]
\centering%
\begin{tabular}{|c||c|cc|cc|cc||c|cc|}
\hline
&  & \multicolumn{2}{c|}{$\left\{ \bar{c}\bar{c}\right\} _{1}$} & 
\multicolumn{2}{c|}{$\left\{ \bar{b}\bar{b}\right\} _{1}$} & 
\multicolumn{2}{c||}{$\left\{ \bar{c}\bar{b}\right\} _{1}$} &  & 
\multicolumn{2}{c|}{$\left[ \bar{c}\bar{b}\right] _{0}$} \\ 
& $J^{P}$ & {chan.} & {thr.} & {chan.} & {thr.} & chan. & {thr.} & $J^{P}$ & 
chan. & {thr.} \\ \hline \hline
$\Lambda _{Q}$ & $1^{+}$ & $\bar{D}^{0}D^{\ast -}$ & 3875 & $B^{+}B^{\ast 0}$ & 
10\thinspace 604 & $\bar{D}^{0}B^{\ast 0}$ & 7190 & $0^{+}$ & $\bar{D}^{0}B^{0}$ & 7144
\\ 
\hline
$\Xi _{Q}$ & $1^{+}$ & $\bar{D}^{\ast 0}D_s^{-}$ & 3976 &$B^{*+} \bar{B}_s^{0}$&10\,692  &$\bar{D}^{*0}\bar{B}_s^{0}$  &7281  & $0^{+}$ &$\bar{D}^0 \bar{B}_s^0 $ & 7232 
\\ \hline \hline
& $0^{+}$ &$D^0 D^{0}$ &$3730$ &$B^+ B^+$  &10\,559  & $\bar{D}^0 B^+$  &7144  & $\cdots $ & $\cdots $ & $\cdots $ \\ 
$\Sigma _{Q}$ & $1^{+}$ &  $D^0 D^{*0}$ &3872   &$B^+ B^{*+}$  &10\,604  &$\bar{D}^{*0} B^+ $  & 7286 & $1^{+}$ &$\bar{D}^{*0} B^+ $  & 7286 \\ 
& $2^{+}$ &$D^{*0} D^{*0}$  &4014  & $B^{*+} B^{*+}$ &10\,649  &$ \bar{D}^{*0} B^{*+} $ &  7332& $\cdots $ & $\cdots $ & $\cdots $ \\ \hline
& $0^{+}$ &$\bar{D}^{0}D_s^{-}$   &3834 &$B^{+} \bar{B}_s^{0}$  &10\,646  & $\bar{D}^0\bar{B}_s^{0}$ &7232  & $\cdots $ & $\cdots $ & $\cdots $ \\ 
$\Xi _{Q}^{\prime }$ & $1^{+}$ & $\bar{D}^{\ast 0}D_s^{-}$ & 3976 &$B^{+} \bar{B}_s^{*0}$&10\,692  &$\bar{D}^0\bar{B}_s^{* 0}$  &7281  & $1^{+}$ &$\bar{D}^0 \bar{B}_s^{*0} $ & 7281    \\ 
& $2^{+}$ &$\bar{D}^{\ast 0}D_s^{* -}$  & 4120 & $B^{*+} \bar{B}_s^{*0}$ & 10\,741 & $\bar{D}^{*0}\bar{B}_s^{* 0}$  & 7423 & $\cdots $ & $\cdots $ & $\cdots $ \\ \hline
& $0^{+}$ &$D_s^{-} D_s^{-}$  &3938  &$\bar{B}_s^0 \bar{B}_s^0$   & 10\,734 &$D_s^{-} \bar{B}_s^{0}$  & 7336 & $\cdots $ & $\cdots $ & $\cdots $ \\ 
$\Omega _{Q}$ & $1^{+}$ & $D_s^{-} D_s^{*-}$  &4082  & $\bar{B}_s^0 \bar{B}_s^{*0}$ &10\,783  &$D_s^{*-} \bar{B}_s^{0}$  &7480  &
 $1^{+}$ &$D_s^{*-} \bar{B}_s^{0}$  & 7480 \\ 
& $2^{+}$ & $D_s^{*-} D_s^{*-}$ & 4226 & $\bar{B}_s^{*0} \bar{B}_s^{*0}$  & 10\,832 & $D_s^{*-} \bar{B}_s^{*0}$  & 7529 & $\cdots $ & $\cdots $ & $\cdots $ \\ \hline
\end{tabular}%
\caption{Thresholds for tetraquark decays in MeV. First column shows the
baryon entering the mass formulas (\ref{eq:tetra3bar}) and (\ref{eq:tetra6}), 
which specifies the tetraquark SU(3)
representation. Next columns indicate pertinent diquarks and their spin.
If more than one decay channel is possible, only the one
with the lowest mass is shown. }
\label{tab:thresholds}
\end{table}
\end{center}
\end{widetext}
\renewcommand{\arraystretch}{1.0} 

There is one important practical reason to use the Cornell potential in the present
context. For a $Q_1{\bar{Q}_2}$ system in color singlet $C_{\rm color}=C_F=4/3$.
In order to compute diquark masses $Q_1Q_2$ (or $\bar{Q}_1\bar{Q}_2$) one has to couple 
quark
color charges to $\bar{\bf 3}$ (or ${\bf 3}$), and then the color factor is $C_{\rm color}=C_F/2=2/3$
 (see e.g. Table III in
Ref.~\cite{Karliner:2014gca}). As this is quite obvious for the Coulomb and spin term,
lattice calculations suggest the same behavior of the confining part
\cite{Nakamura:2005hk}.

Therefore, once the potential parameters are fixed from the $c\bar{c}$ and $b\bar{b}$ meson spectra
we can compute diquark masses by rescaling the color factors and the string tension in (\ref{eq:Cornell})
by a factor of 2.

\begin{figure}[h!]
\centering
\includegraphics[width=7cm]{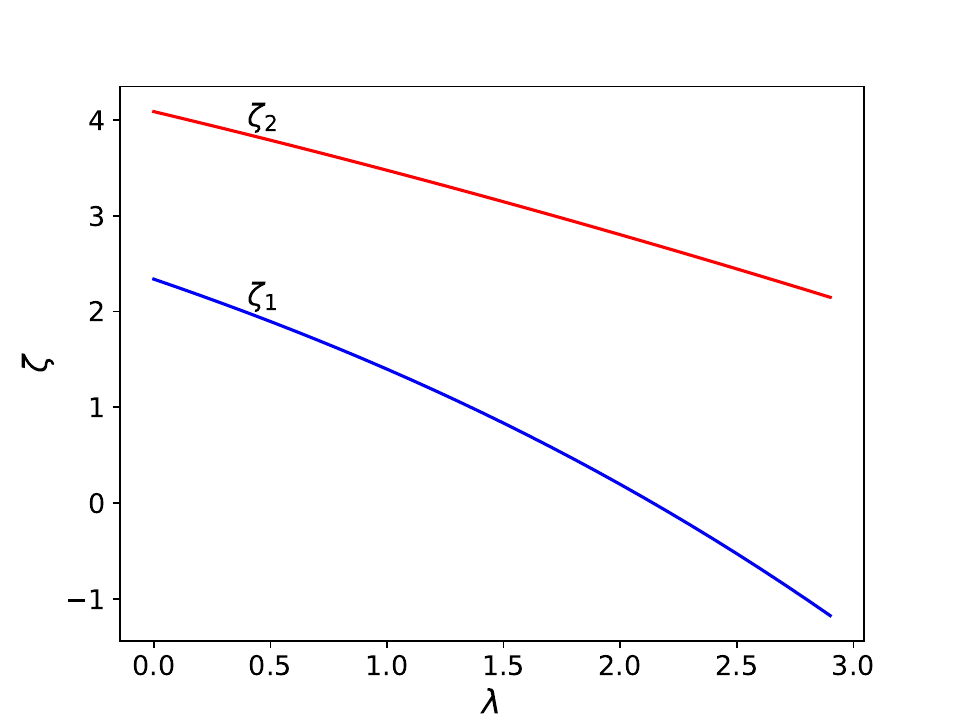} \\
\includegraphics[width=7cm]{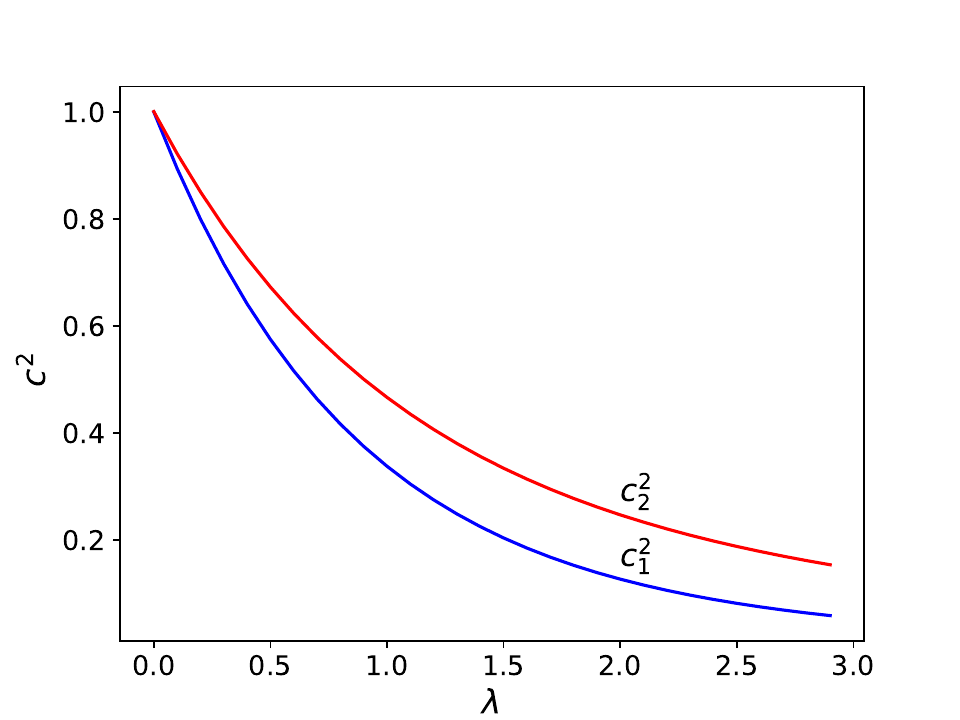}
\vspace{-0.2cm}%
\caption{Dimensionless energies $\zeta_n$ and normalization factors $c_n^2$  for the ground
and sate and first excited state as functions of of  $\lambda$.
}%
\label{fig:zetanorm}%
\end{figure}

We are looking for a
solution of the Schr{\"o}dinger equation in terms of a function $u_{n}(r)$ defined as follows
\begin{equation}
\psi_{nl=0 m=0}(r,\theta,\varphi)=R_{0}^{n}(r)Y_{00}(\theta,\varphi)= \frac
{u_{n}(r)}{r} \frac{1}{\sqrt{4\pi}}.
\end{equation}
It is convenient to introduce a dimensionless variable $\rho$
\begin{equation}
r=\left(  \frac{1}{2 \sigma \mu }\right)  ^{1/3} \rho
\end{equation}
and rescaled dimensionless parameters $\lambda$ and $\zeta$:
\begin{equation}
\lambda  =\left(  \frac{2 \mu}{\sigma^{1/2}}\right)  ^{2/3}%
\kappa,~~~
\zeta =\left(  \frac{2 \mu}{\sigma^{2}}\right)  ^{1/3}%
E.\label{eq:newpars}%
\end{equation}
With these substitutions the Schr{\"o}dinger equation takes a very simple
form
\begin{equation}
u^{\prime\prime}+\left[  \frac{\lambda}{\rho}-\rho+\zeta\right]  u
=0 \, .\label{eq:fullu}%
\end{equation}
The results for the rescaled energies $\zeta_i$ are shown in upper panel of Fig~\ref{fig:zetanorm}.
We choose normalization
\begin{equation}
\int\limits_{0}^{\infty }d\rho \left\vert u(\rho )\right\vert ^{2}=1 \, .
\end{equation}%

Now, we need to compute the hyper-fine splitting. In the first order of perturbation theory for $l=0$ states we have
\begin{equation}
\Delta _{\text{hf}}^{(s)}E_{n}=\frac{2}{3}C_{F}\alpha _{\text{s}}\frac{%
2\sigma }{m_{1}+m_{2}}\left( \left. \frac{u_{n}(\rho )}{\rho }\right\vert
_{\rho =0}\right) ^{2}(s_{1}\cdot s_{2}) \, .
\label{eq:Deltahf}
\end{equation}

In Ref.~\cite{Praszalowicz:2022sqx} we have solved Eq.~(\ref{eq:fullu}) semi-analytically treating the Coulmb part
as a perturbation, since for $\lambda=0$ Eq.~(\ref{eq:fullu}) reduces to the Airy equation.
While this method is quite accurate as far as the eigenvalues $\zeta_n$
are concerned, it fails for the hyper-fine splitting (\ref{eq:Deltahf}) where the value of the wave function
in the origin is needed. Therefore, here we have decided to solve Eq.~(\ref{eq:fullu})
numerically. Because for $l=0$ function $R(r)$ is
constant at $r=0$, function $u_n(\rho )=c_n\rho +\mathcal{O}(\rho ^{2})$ for
small $\rho $.
In Fig~\ref{fig:zetanorm} we plot normalization constants $c_n^2$ for $n=1$ and $2$.
 As a result  the mass of the $Q_{1}\bar{Q}_{2}$ meson (and its antiparticle)
of spin $s$
reads as follows
\begin{align}
(M_{Q_{1}\bar{Q}_{2}})_{n}^{s} & =m_{1}+m_{2}+\left( \frac{\sigma ^{2}}{2\mu }\right) ^{1/3}\zeta_{n}  
\label{eq:Mesonmass} \\
& +\frac{2}{3}C_{F}\alpha _{\text{s}}\frac{2\sigma }{m_{1}+m_{2}} c_n^2
\left\{ 
\begin{array}{ccc}
-3/4 & \text{for} & s=0 \\ 
&  &  \\ 
+1/4 & \text{for} & s=1%
\end{array}%
\right. \, . \notag
\end{align}

As explained earlier, diquark masses can be computed from the same formula 
by rescaling $C_F \rightarrow C_F/2$ and $\sigma \rightarrow \sigma/2$. This
rescaling changes the value of the parameter 
\begin{equation}
\lambda \rightarrow \lambda^{\prime}=\lambda/4^{1/3} \, .
\label{eq:lamprim}
\end{equation}

Note that actual value of $\lambda$ in Eq.~(\ref{eq:fullu}) depends on the system considered, as it depends on $\mu$,
both explicitly (\ref{eq:newpars}) and implicitly, since also $\kappa$ is a function of $\mu$. 
For this new value  $\lambda'$ we have different energies $\zeta'_n$ and 
new wave functions
leading to a new value of $c_n \rightarrow c'_n$. Final mass formula for
a diquark is therefore given as follows
\begin{align}
(M_{\bar{Q}_1\bar{Q}_2})_{n}^{s} & =m_{1}+m_{2}+\left( \frac{\sigma ^{2}}{8\mu }\right) ^{1/3}\zeta'_{n}  
\label{eq:diqmass} \\
& +\frac{1}{3}C_{F}\alpha _{\text{s}}\frac{\sigma }{m_{1}+m_{2}} {c_n^{\prime \, 2}}
\left\{ 
\begin{array}{ccc}
-3/4 & \text{for} & s=0 \\ 
&  &  \\ 
+1/4 & \text{for} & s=1%
\end{array}%
\right. \, . \notag
\end{align}

Note that for identical quarks $s=0$ configuration is Pauli forbidden. In practice
we shall consider only two lowest states: the ground state $n=1$ and the first
radially excited state $n=2$.

\subsection{Fitting procedure}
\label{ssec:fits}

As the first step we will use Eq.~(\ref{eq:Mesonmass}) to fix potential parameters from $n=1$
states shown in Table~\ref{tab:onia}. We have decided to perform our own dedicated
 fits, rather than use the global fits to all known quarkonia states. This is because we
 are interested only in the ground states both for mesons and diquarks, however we will see
 that $n=2$
 excited states are quite well reproduced within the accuracy of the present approach.

\begin{table}[h!]
\begin{tabular}{|c|cr|cr|}
\hline
$(n,s)$ &  &  MeV&  & MeV \\ 
\hline
(1,0) & $\eta _{c}(1S_0)$ & 2984 & $\eta _{b}(1S_0)$ & 9399 \\ 
(1,1) & $J/\psi (1S_1) $& 3097 & $\Upsilon (1S_1)$ & 9460 \\ 
(2,0) & $\eta _{c}(2S_0) $ & 3637 &$ \eta _{b}(2S_0)$ & 9999 \\ 
(2,1) &$ \psi (2S_1)$ & 3686 & $\Upsilon (2S_1) $ & 10023 \\
\hline
\end{tabular}%
\caption{Lowest quarkonia states used in the fits of the Cornell potential parameters.}
\label{tab:onia}
\end{table}%

From Table \ref{tab:onia} we find average $n=1$ masses:
\begin{eqnarray}
\overline{M}_{c\bar{c}} =\frac{3M_{J/\psi }+M_{\eta _{c}}}{4} &=& 3.069~{\rm GeV}\, ,  \notag \\
\overline{M}_{b\bar{b}} =\frac{3M_{\Upsilon }+M_{\eta _{b}}}{4}&=&9.445~{\rm GeV}
\end{eqnarray}%
and average $n=1$ spin splittings
\begin{equation}
\delta _{\text{hf}}E(c\bar{c}) =113~{\rm MeV},~~
\delta _{\text{hf}}E(b\bar{b}) =61~{\rm MeV} \, .
\end{equation}

We  first solve numerically Eq.\,(\ref{eq:fullu}) for $0 \le \lambda \le 3$ and tabulate
energy levels $\zeta_{1,2}(\lambda)$ and constants $c_{1,2}(\lambda)$. 
The results are shown in Fig~\ref{fig:zetanorm}.
We have checked our
numerical results comparing with two semi-analytical solutions: one when we solve the Airy equation
and treat the Coulomb part as a perturbation and second, when we solve the Coulomb part and treat the
confining part as a perturbation.

\begin{figure}[h!]
\centering
\includegraphics[width=7cm]{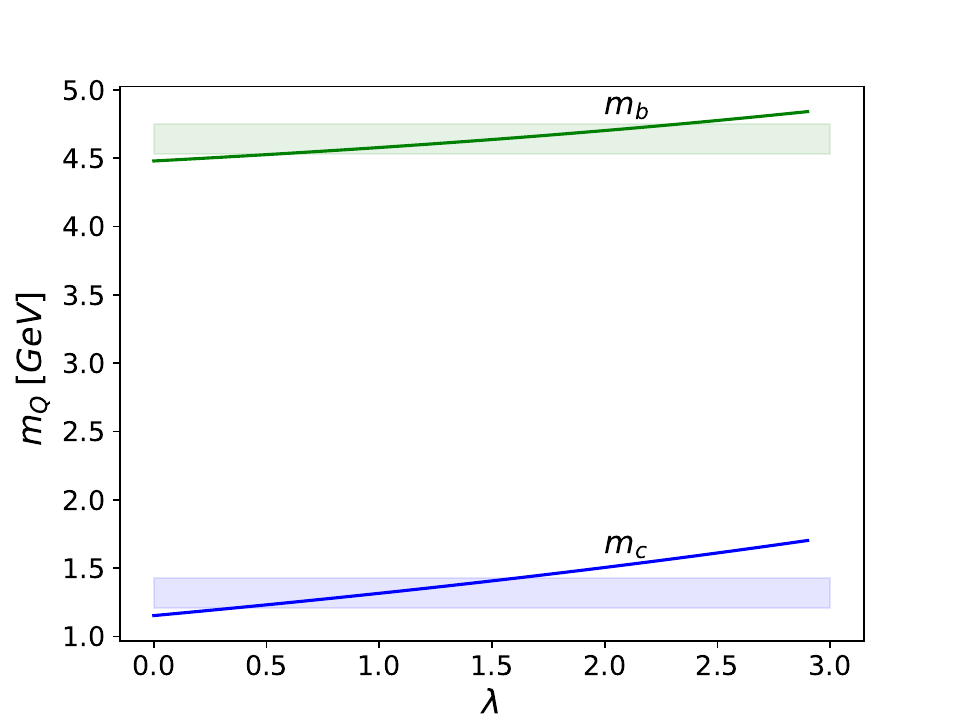} 
\vspace{-0.2cm}%
\caption{Charm (lower blue line) and bottom (upper green line) quark masses in MeV 
as functions of $\lambda$
extracted from Eq.~(\ref{eq:mQfix})
for $\sigma=0.2$~GeV$^2$. Shaded areas
correspond to Eq.\,(\ref{eq:mQbaryon}).}
\label{fig:mQvslam}%
\end{figure}

Next, we fix $\sigma$ and find $c$ and $b$ quark masses as functions of $\lambda$ from the
average ground state masses:
\begin{equation}
\overline{M}_{Q\bar{Q}}=2m_{Q}+\left( \frac{\sigma ^{2}}{m_Q }\right) ^{1/3}\zeta_{1}(\lambda) \, .
\label{eq:mQfix}
\end{equation}
The result is plotted in Fig.~\ref{fig:mQvslam} for $\sigma=0.2$~GeV$^2$. We see
rather moderate dependence of heavy quark masses on $\lambda$. Shaded areas
show mass limits of Eq.\,(\ref{eq:mQbaryon}) that follow from the heavy baryon 
phenomenology in the present approach \cite{Praszalowicz:2022sqx}. We see that heavy quark masses extracted
from charmonia or heavy baryons are compatible, which proves the consistency of our approach.

Then, from the hyperfine splitting we find the value of $\alpha_s(m_Q)$ 
\begin{equation}
\delta _{\rm hf}E_{1}(Q\bar{Q})=\frac{2}{3}C_{F}\alpha _{\rm s}(m_Q)\frac{\sigma }{m_{Q}} c_1^2(\lambda)
\label{eq:alphafix}
\end{equation}
as a function of $\lambda$.

\begin{figure}[h!]
\centering
\includegraphics[width=7cm]{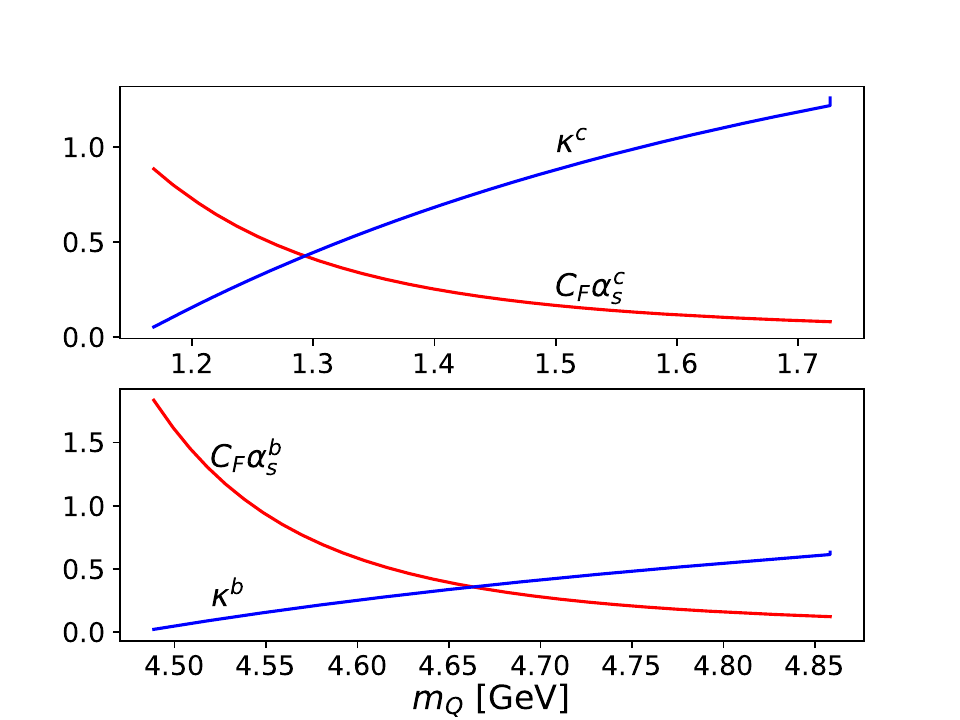} 
\vspace{-0.2cm}%
\caption{Dependence of $C_F \alpha_s$ (red) and $\kappa$ (blue) on the heavy quark mass
for $\sigma=0.2$ GeV$^2$,
charm -- upper panel, bottom -- lower panel. The point when the two lines cross
corresponds to the model heavy quark mass for given string tension $\sigma$.}%
\label{fig:kappalpha}%
\end{figure}

Since for a given $\lambda$ the quark mass $m_Q$ is fixed by Eq.\,(\ref{eq:mQfix}) we can 
compute $\kappa_Q(\lambda)$ both for charm and bottom
from Eq.\,(\ref{eq:newpars}). However,  $\kappa_Q(\lambda)=C_F \alpha_s(m_Q,\lambda)$,
therefore
we can find $\lambda_Q^{\rm sol}$ for which this equality is satisfied. Since there is one-to-one correspondence
between $\lambda_Q$ and $m_Q$ (see Fig.~\ref{fig:mQvslam}),
in Fig.~\ref{fig:kappalpha} we plot $\kappa$ and $C_F \alpha_s$ 
 in terms of the corresponding charm  (top panel) and bottom (bottom panel) mass
for $\sigma=0.2$ GeV$^2$. Two lines cross at the quark mass corresponding to $\lambda_Q^{\rm sol}$.

\begin{figure}[h!]
\centering
\includegraphics[width=7cm]{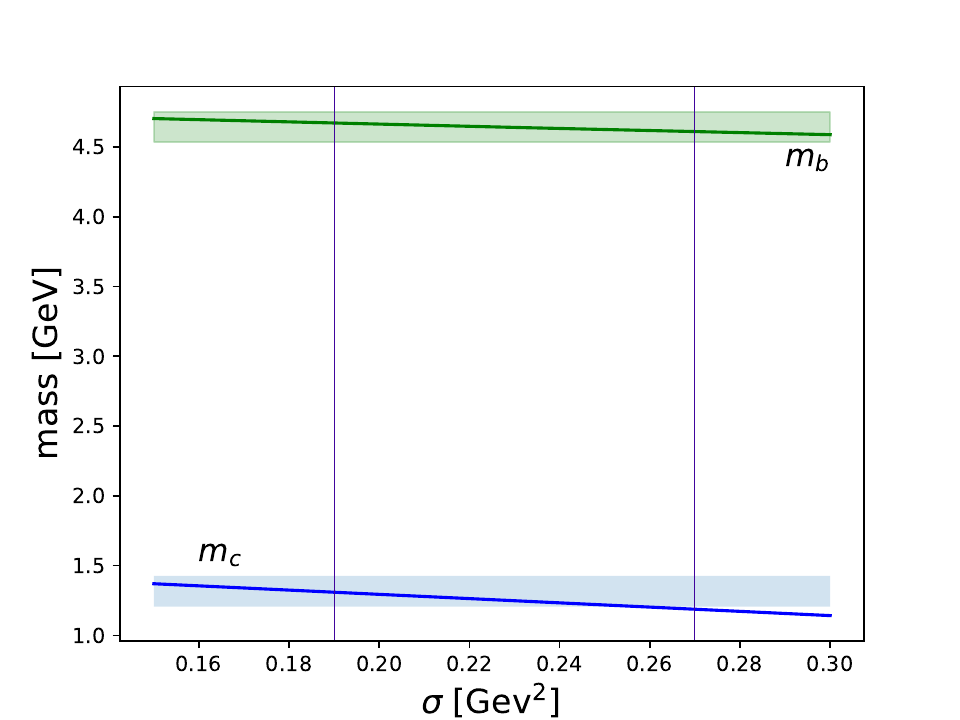} 
\vspace{-0.2cm}%
\caption{Charm (lower blue line) and bottom (upper green line) quark masses in GeV 
as functions of $\sigma$ obtained from the fits to $1S$ states. Shaded areas
correspond to Eq.\,(\ref{eq:mQbaryon}). Vertical lines correspond to the best fits to $2S$ states:
left  -- charm, right -- bottom, see Sect.~\ref{sec:quarkonia}.}
\label{fig:mQvssig}%
\end{figure}

\begin{figure}[h!]
\centering
\includegraphics[width=7cm]{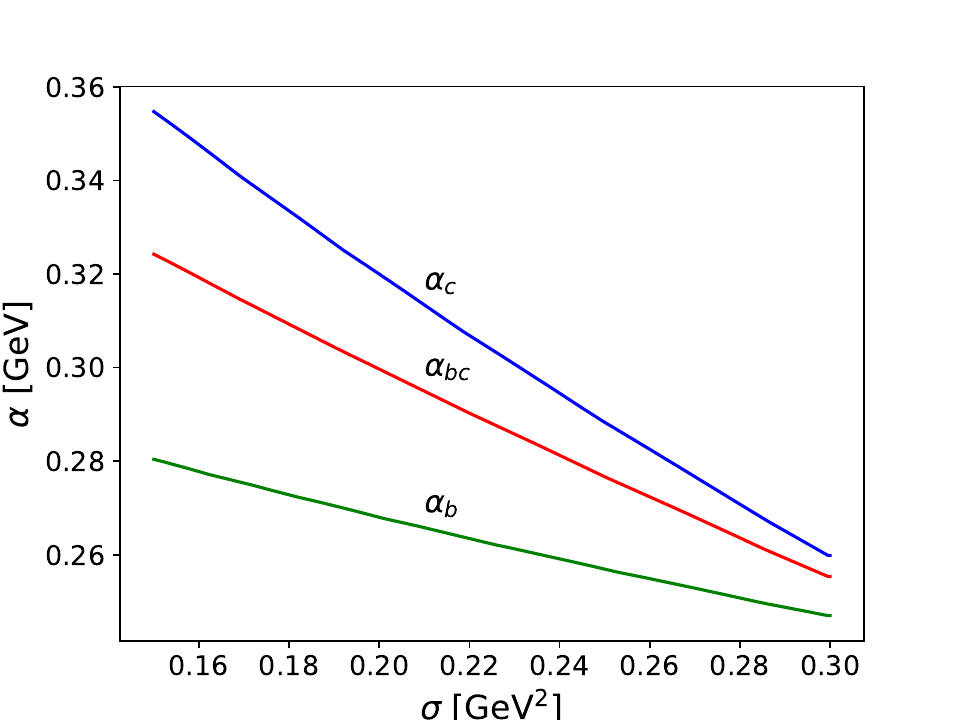} 
\vspace{-0.2cm}%
\caption{Strong coupling constants $\alpha_s(2 \mu)$ for charm (upper blue line), bottom (lower green line) 
and for $\mu$ equal to the reduced mass of the $bc$ system (middle red line, see Sect.~\ref{sec:quarkonia})
as functions of the string tension $\sigma$. }
\label{fig:alvssig}%
\end{figure}

In this way, for given $\sigma$ we find unique values of $m_{c,b}(\sigma)$ and $\alpha_s(m_{c,b}(\sigma))$
that fit $1S$ ground state quarkonia masses. The results are plotted in Figs.~\ref{fig:mQvssig} and \ref{fig:alvssig}.
We see that quark mass dependence on $\sigma$ is relatively weak, and that masses extracted from $Q\bar{Q}$ mesons fall within
the  range (\ref{eq:mQbaryon}) corresponding to the baryonic fits. This proves the consistency of our approach
that combines the soliton model with the non-relativistic theory of heavy quark bound states. Nevertheless, 
the $m_b-m_c$ mass difference changes in this $\sigma$ range by about 100~MeV, which -- as we will see -- is a source
of uncertainty in the determination of the $cb$ tetraquark mass.

\subsection{Numerical results for quarkonia}
\label{sec:quarkonia}

\begin{figure}[h]
\centering
\includegraphics[width=6.5cm]{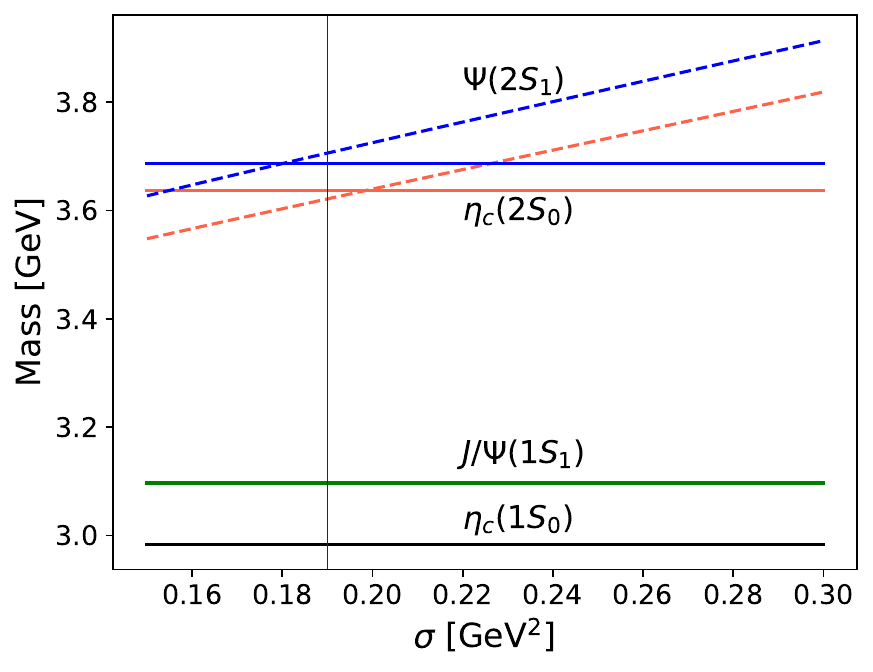} \\
\includegraphics[width=6.5cm]{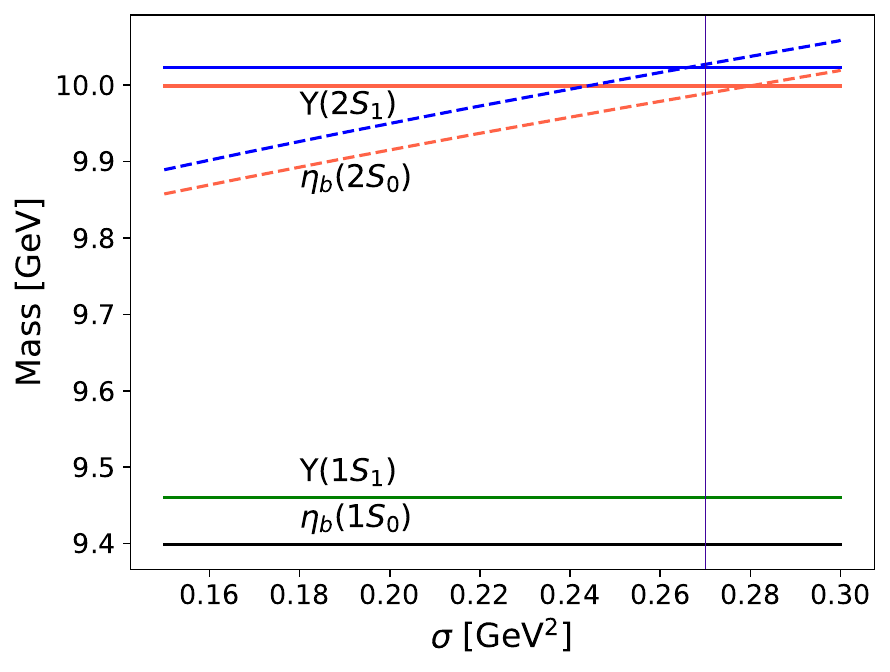}
\vspace{-0.2cm}%
\caption{Masses of lowest $S$ state charmonia (upper panel) and bottomia (lower panel)
listed in Tab.~\ref{tab:onia}. Dashed lines correspond to the fits described in the text, 
horizontal lines to the experimental values. $1S$ states are used as input.
Vertical lines indicate the values
of $\sigma$ for which  $2S$ mesons are best reproduced.}%
\label{fig:onia}%
\end{figure}

Since all parameters are now fixed from the ground states, excited state
masses are predictions. Results are plotted in Fig.~\ref{fig:onia}. We see
that first excited states in the charm and bottom sector cannot be fitted 
by the same value of $\sigma$. The best fit for charmonia
requires $\sigma_c \simeq 0.19$~GeV$^2$, while in the bottom sector
$\sigma_b \simeq 0.27$~GeV$^2$. This agrees quite well with the results 
of global fits of Ref.~\cite{Mateu:2018zym}, which give $0.164\pm 0.011$~GeV$^2$ 
and $0.207\pm 0.011$~GeV$^2$ respectively. Still, the error on excited charmonia
masses at $\sigma=\sigma_b$ or bottomia masses at $\sigma=\sigma_c$ is of the 
order of 70 MeV, {\em i.e.} 2\% in the case of charmonia and less than 1\% for bottomia.
We therefore restrict the range of the string tension to 
\begin{equation}
0.19~{\rm GeV}^2 \le \sigma \le 0.27~{\rm GeV}^2\, ,
\label{eq:siglimits}
\end{equation}
which in terms of the quark masses corresponds to:
\begin{eqnarray}
1.19~{\rm GeV} \le &m_c& \le 1.31~{\rm GeV}, \nonumber \\
4.61~{\rm GeV} \le &m_b& \le 4.67~{\rm GeV}\, ,
\label{eq:qlimits}
\end{eqnarray}
which narrows the allowed range (\ref{eq:mQbaryon}) following from the fits to heavy baryons.

We should stress once again that the above result is by no means trivial. Quark masses obtained from baryon spectra 
could in principle differ from dynamical inference from the meson sector. The fact that both sectors are compatible
reinforces confidence in the consistency of the current approach.

\begin{figure}[h!]
\centering
\includegraphics[width=6.4cm]{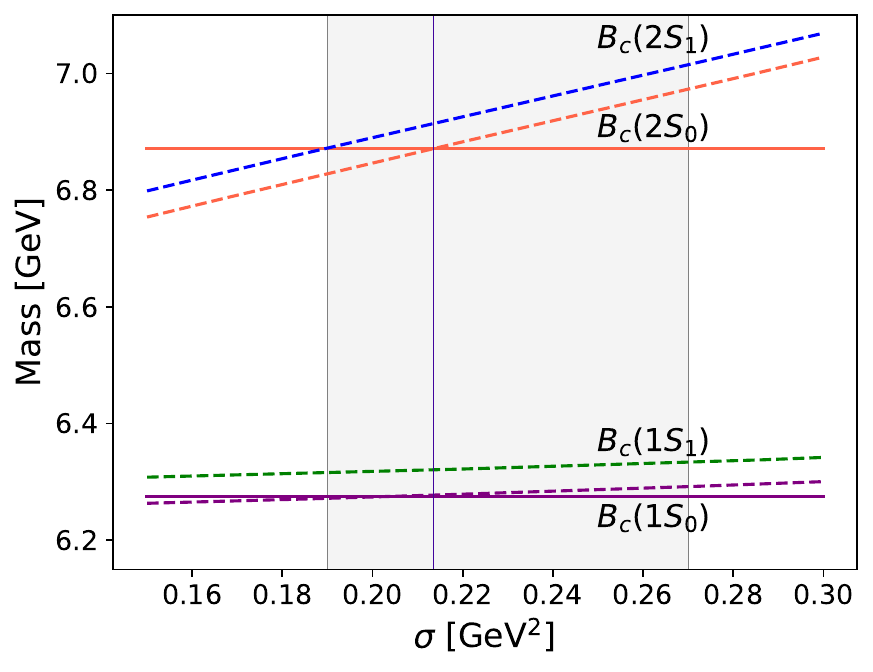} 
\vspace{-0.2cm}%
\caption{Predicted masses of $B_c$ mesons as functions of the string tension $\sigma$ shown as dashed lines.
Solid lines denote two known spin $s=0$ mesons $B_{c}^{+}(1S_0,6274.5)$ and
$B_{c}^{\pm}(2S_0,6871.2)$~\cite{Workman:2022ynf}. Vertical line indicates the value
of $\sigma$ for which both $B_c$ mesons are best reproduced. Shaded area corresponds
to the limits of Eq.~(\ref{eq:siglimits}).}
\label{fig:Mbcmeson}%
\end{figure}

We can now easily predict masses of $c{\bar{b}}$ or $\bar{c}b$ mesons, two of which, namely spin zero
$B_{c}^{+}(1S_0,6274.5)$ and
$B_{c}^{\pm}(2S_0,6871.2)$ mesons are listed in the PDG~\cite{Workman:2022ynf}. To this end we need
to estimate the value of $\alpha_s(2 \mu)$, where $\mu$ is the reduced mass (\ref{eq:redmass})
of the $cb$ system. To this end we use the evolution formula
\begin{equation}
\alpha _{s}(m_{b})=\frac{\alpha _{s}(m_{c})}{1+\frac{\beta _{0}}{2\pi }%
\alpha _{s}(m_{c})\ln (m_{b}/m_{c})},
\end{equation}%
which allows to compute model $\beta_0$ as a function of the string tension\footnote{Remember that quark masses
are in one-to-one correspondence with the string tension.}
\begin{equation}
\beta _{0}=2\pi \frac{1/\alpha _{s}(m_{b})-1/\alpha _{s}(m_{c})}{\ln
(m_{b}/m_{c})}.
\end{equation}
From this we obtain $\alpha_s(2\mu)$, which is plotted
as a red line in Fig.~\ref{fig:alvssig}. The resulting masses are shown in Fig.~\ref{fig:Mbcmeson}.
For known spin $s=0$ mesons we have:
\begin{eqnarray}
6.26~{\rm GeV} \le &m(B_c(1S_0,6.275))& \le  6.28~{\rm GeV} \, , \nonumber \\
6.83~{\rm GeV} \le &m(B_c(2S_0,6.871))& \le  6.97~{\rm GeV} \, ,
\label{eq:Bcmasses}
\end{eqnarray}
where the limits correspond to (\ref{eq:siglimits}). We also predict for spin $s=1$ states
\begin{eqnarray}
6.32~{\rm GeV} \le &m(B_c(1S_1))& \le  6.34~{\rm GeV} \, , \nonumber \\
6.87~{\rm GeV} \le &m(B_c(2S_1))& \le  7.02~{\rm GeV} \, .
\label{eq:Bcmasses1}
\end{eqnarray}
The best fit, shown by vertical line in Fig.~\ref{fig:Mbcmeson}, is for $\sigma=0.21$~GeV$^2$ giving for spin $s=1$ 
$M_{b\bar{c}}(1S_1)=6.32$~GeV and $M_{b\bar{c}}(2S_1)=6.91$~GeV. For the most recent survey of
$B_c$ states see Ref.~\cite{Li:2023wgq}.

Summarizing: we have fixed Cornell potential parameters from the $c\bar{c}$ charmonia and $b \bar{b}$ bottomia spectra
and computed without any further inputs masses of the ground state $c\bar{b}$ (or $\bar{c}b$) mesons that for the experimentally
measured states agree very well with data.

\subsection{Numerical results for diquarks}
\label{ssec:numQQ}

\begin{figure}[h!]
\centering
\includegraphics[width=7cm]{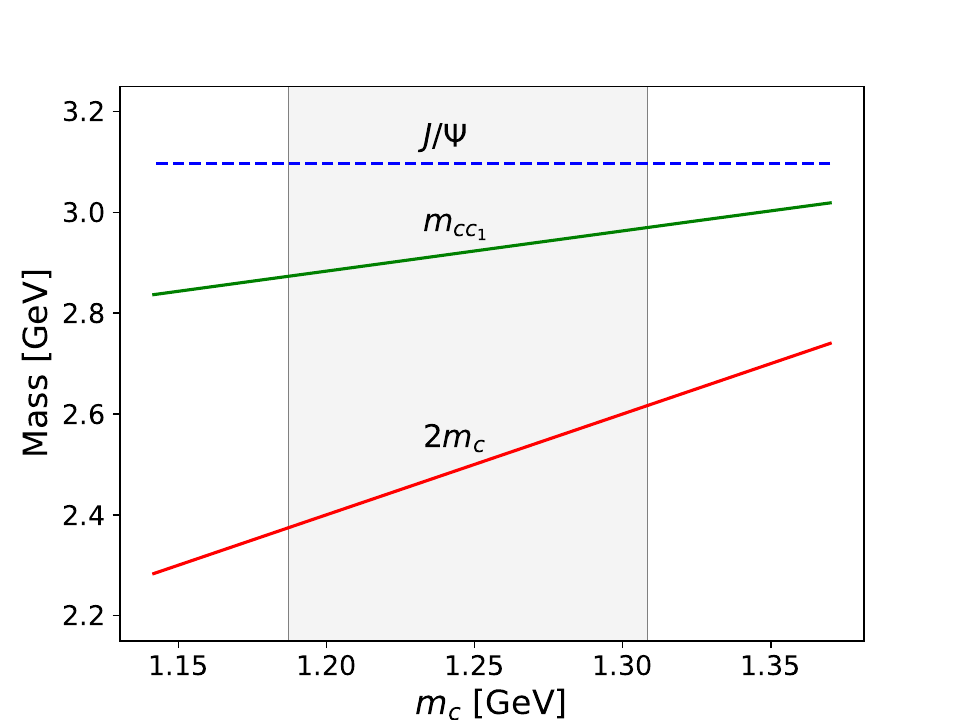} \\
\includegraphics[width=7cm]{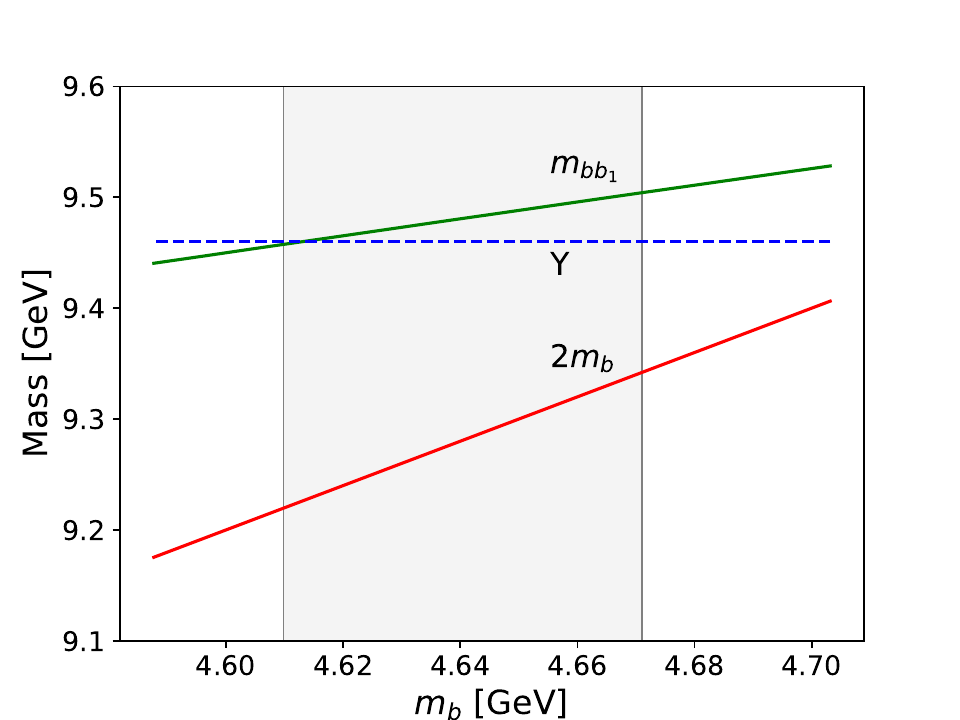} \\
\vspace{-0.2cm}%
\caption{Spin $s=1$ charm (upper panel) and bottom (lower panel) diquark masses in GeV  (green solid lines)
as functions of $m_{c,b}$.
Horizontal dashed lines show $J/\Psi$  and $\Upsilon$ masses respectively, red solid lines correspond to
$2 m_{c,b}$. Vertical lines indicate the values of  $m_{c,b}$ corresponding to
 $\sigma$ for which  $2S$ mesons are best reproduced.  
Shaded area corresponds
to the limits of Eq.~(\ref{eq:qlimits}).
}
\label{fig:diqccbb}%
\end{figure}

Having constrained the parameters of the Cornell potential, we can now -- with the help of Eqs.~(\ref{eq:diqmass}) and (\ref{eq:lamprim}) --
compute the diquark masses. In Fig.~\ref{fig:diqccbb} we plot $cc$  and $bb$ spin $s=1$ diquark masses as functions of $m_{c,b}$ rather than
$\sigma$. The results are very similar to the ones obtained previously in Ref.~\cite{Praszalowicz:2022sqx}, with one difference. Namely, the slope
of diquark masses obtained here is smaller than 1 (with respect to $m_{Q}$), whereas in Ref.~\cite{Praszalowicz:2022sqx}
the slope was slightly larger than 1.
This means that the tetraquark masses, which are proportional to $m_{\bar{Q}\bar{Q}}-m_Q$, decrease with $m_Q$, while in Ref.~\cite{Praszalowicz:2022sqx}
they were increasing as functions of the heavy quark mass. Numerically, however, the results are very similar and show slower
increase than the total mass of their constituents.

\begin{figure}[h!]
\centering
\includegraphics[width=7cm]{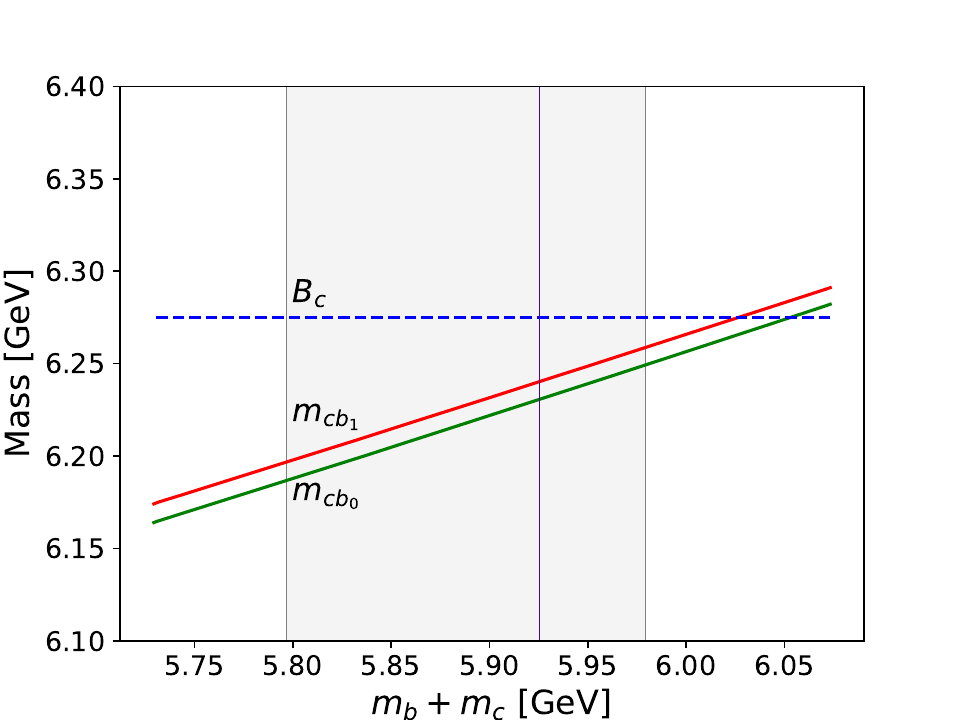} 
\vspace{-0.2cm}%
\caption{Masses of $cb$ diquarks  (solid lines)
as functions of $m_{c}+m_b$. Lower line corresponds to $s=0$ and upper one
to $s=1$.
Horizontal dashed line shows $B_c(1S)$  mass. Vertical line indicates the value
of $m_b+m_c$ corresponding to $\sigma$ for which  $B_c$ mesons are best reproduced.  
Shaded area corresponds
to the limits of Eq.~(\ref{eq:qlimits}).}
\label{fig:diqcb}%
\end{figure}

In Fig.~\ref{fig:diqcb} we plot $cb$ diquark masses both for spin 0 and spin 1 as functions
of $m_c+m_b$. We see that, similarly to $m_{cc}$, the diquark mass is smaller than the 
relevant meson mass. In the case of $m_{bb}$ the diquark mass is larger than the mass of
$\Upsilon$.

\section{Tetraquark Masses}

\label{sec:masses}

\subsection{Anti-triplet masses}

\begin{center}
\begin{figure}[h]
\centering
\includegraphics[height=6.3cm]{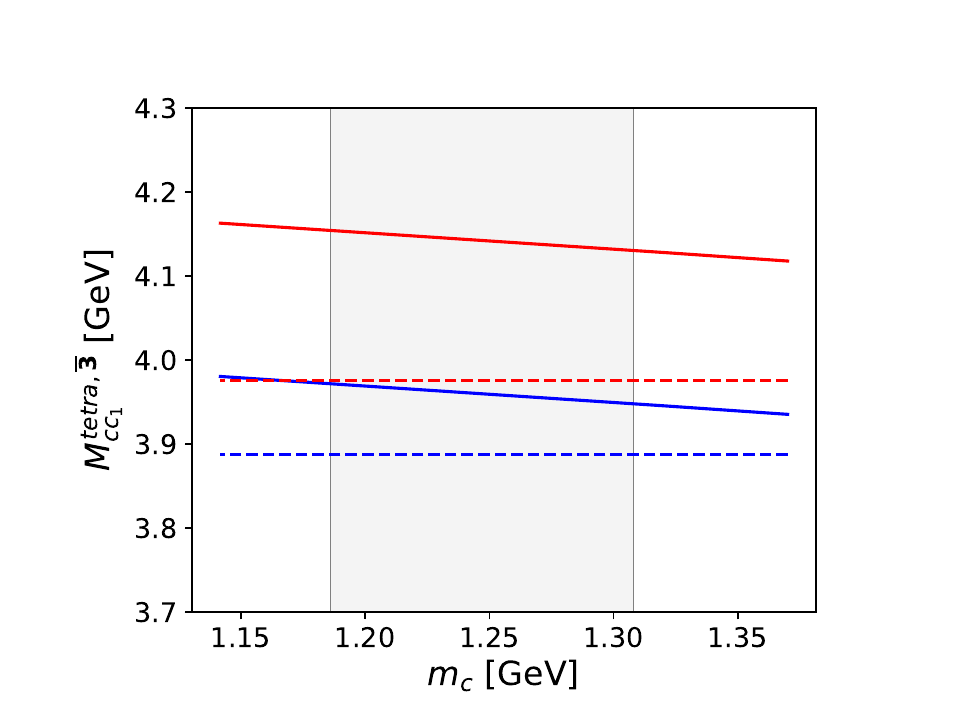} \\
\includegraphics[height=6.3cm]{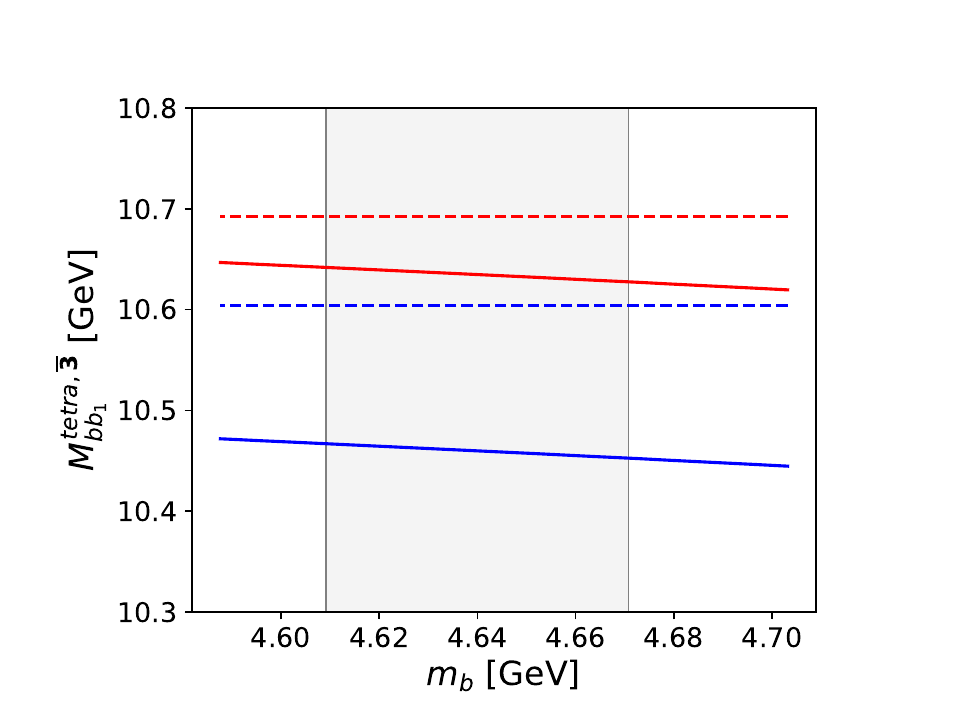} 
\caption{The lightest
non-strange (solid blue, bottom) and strange (solid red, top) anti-triplet tetraquark masses (charm, upper panel; bottom, lower panel)
as  functions of the heavy quark mass. Horizontal dashed lines correspond to the pertinent thresholds
(non-strange, bottom;  strange, top)
given in Table \ref{tab:thresholds}. Shaded areas correspond the heavy
quark mass ranges (\ref{eq:qlimits}). Vertical lines indicate the values of  $m_{c,b}$ corresponding to
 $\sigma$ for which  $2S$ mesons are best reproduced.  
 }
  \label{fig:3barmass}%
\end{figure}
\end{center}

To compute  tetraquark masses in
flavor $\boldsymbol{\overline{3}}$ we shall use Eq.~(\ref{eq:tetra3bar})
and the numerical results for the diquark masses from the previous section.
Since identical quarks have to be in the spin 1 state, anti-triplet tetraquarks are $J^P=1^+$.
The results are plotted in Fig.~\ref{fig:3barmass}. We see that charm
tetraquark masses are above the threshold, while in the case of bottom
there are rather deeply bound states both for non-strange and strange tetraquarks.
The lightest non-strange charm tetraquark is approximately $70 \div 95 $~MeV above the 
threshold, while the strange one is $155 \div 180$~MeV above the threshold. 
On the contrary, bottom tetraquarks are bound by $140 \div 150$~MeV and $50 \div 65 $~MeV
for non-strange and strange tetraquarks, respectively.
These masses
are in agreement with  our previous work \cite{Praszalowicz:2022sqx}, except the
$m_{c,b}$ dependence, which -- as explained in Sect.~ \ref{ssec:numQQ} -- has a different slope.
Our results are also in a very good agreement with
predictions of Ref.~\cite{Eichten:2017ffp}.

\begin{center}
\begin{figure}[h]
\centering
\includegraphics[height=5.2cm]{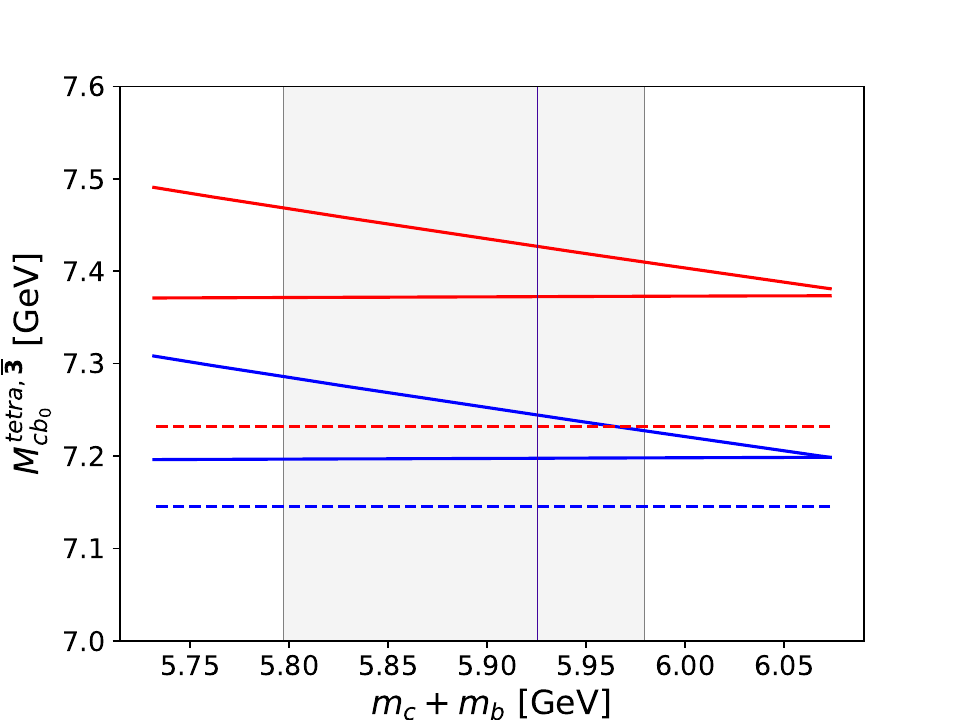} \\
\includegraphics[height=5.2cm]{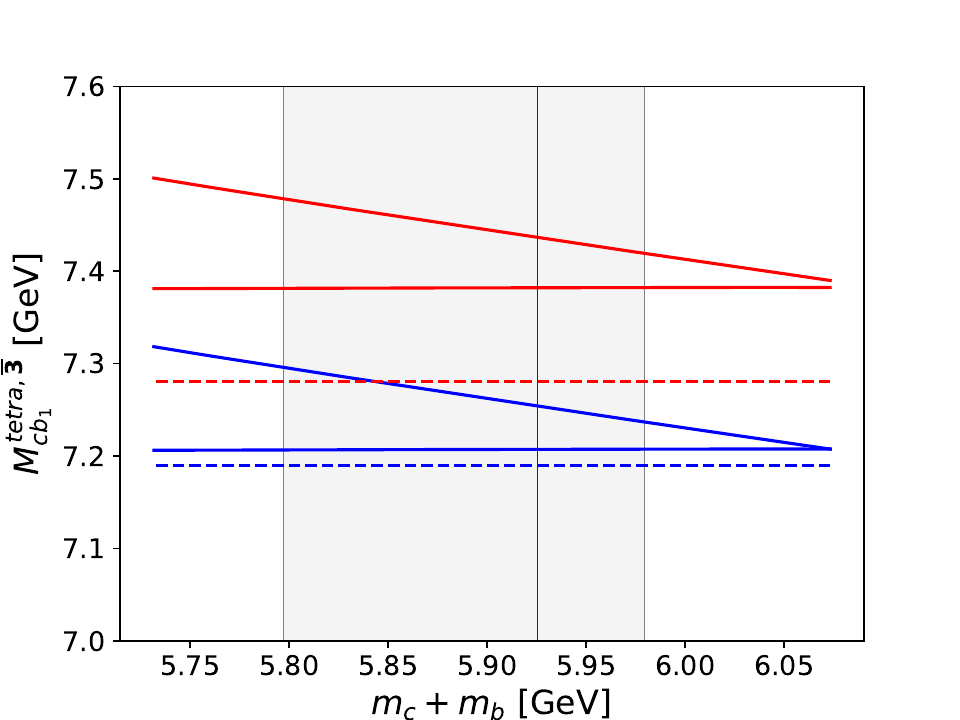} 
\caption{
non-strange (solid blue, bottom) and strange (solid red, top) anti-triplet $cb$ tetraquark masses (spin 0, upper panel; spin 1, lower panel)
as  functions of $m_{c}+m_b$. Upper solid lines correspond to mass computed from the charm sector, while lower ones to the $b$ sector.
Horizontal dashed lines correspond to the pertinent thresholds
(non-strange, bottom;  strange, top)
given in Table \ref{tab:thresholds}. Shaded areas show the heavy
quark mass ranges (\ref{eq:qlimits}). Vertical lines indicate the values of  $m_{c}+m_b$ corresponding to
 $\sigma$ for which  $2S$ mesons are best reproduced.}%
 \label{fig:3barmass1}
\end{figure}
\end{center}

Our new result in the present work are predictions for masses of $cb$ tetraquarks. The results are plotted in Fig.~\ref{fig:3barmass1}.
Here, unlike in the case of identical quarks, both spin configurations of the $cb$ diquarks are possible: spin 0 shown in the upper
panel of Fig.~\ref{fig:3barmass1} and spin 1 in the lower panel. Moreover, we have two sets of predictions based on Eq.~(\ref{eq:tetra3bar}),
where one can choose for $Q$ either $c$ or $b$. In principle both determinations should coincide, we see, however, a difference
of the order of $100 \div 30$~MeV due to the variation of $m_b-m_c$ mass difference with $\sigma$ discussed at the
end of Sect.~\ref{ssec:fits}. The predictions from
the bottom sector are lower and almost independent of the quark masses, while the predictions from the charm sector decrease with $m_c+m_b$.

\begin{table}
\begin{tabular}
[c]{|c|c|cc|c|}%
\hline
\multirow{2}{*}{baryon} & \multirow{2}{*}{$QQ_{S}$} & \multirow{2}{*}{mass} & \multirow{2}{*}{threshold} & $m_{Q}$ or \\
             &                 &           &                 & $m_{c}+m_{b}$\\
 \hline
 $\Lambda_{Q}$ & \multirow{2}{*}{$cc_{1}$} & 3.948 & 3.887 &\multirow{ 2}{*}{1.307}\\
 $\Xi_{Q}$ &  & 4.130 & 3.976 & \\
 \hline
 $\Lambda_{Q}$ &\multirow{2}{*}{ $bb_{1}$ }& {\bf 10.467} & 10.604 & \multirow{ 2}{*}{4.609}\\
 $\Xi_{Q}$ &  & {\bf 10.642} & 10.692 & \\
 \hline
\multirow{2}{*}{ $\Lambda_{Q}$ }& $cb_{0}$ & 7.197--7.241 & 7.145 & \multirow{ 4}{*}{5.932}\\
  & $cb_{1}$ & 7.207--7.251 & 7.190 & \\
  \cline{2-4}
 \multirow{2}{*}{$\Xi_{Q}$} & $cb_{0}$ & 7.327--7.424 & 7.232 & \\
  & $cb_{1}$ & 7.382--7.334 & 7.281 & \\
  \hline
\end{tabular}
\caption{Tetraquark masses in GeV at quark masses  corresponding to
 $\sigma$ for which  $2S$ mesons are best reproduced. Index $S$ refers to the diquark
 spin, which in the $\bar{\bf 3}$ case is equal to the baryon spin. States below the threshold
 are displayed in bold face.}
 \label{tab:masses3bar}
\end{table}

The results of this Section are summarized in Table~\ref{tab:masses3bar} where we quote our predictions
for the tetraquark masses at quark masses corresponding to
 $\sigma$ for which  $2S$ mesons are best reproduced. This means that for each sector
 we have in fact different $\sigma$. It is therefore surprising that the 
 $m_{c}+m_{b}$ mass for $cb$ tetraquarks is practically equal to the sum of $m_c$ and $m_b$ masses determined
 from the $c$ and $b$ sector separately ({\em i.e.} for different $\sigma$).
 
 We see from Table~\ref{tab:masses3bar}  that only $bb$ tetraquarks, both strange and non-strange, 
 are bound confirming results from 
 Refs.~\cite{Praszalowicz:2022sqx,Eichten:2017ffp}. Interestingly $cb$ non-strange tetraquark of spin 1
 is only 17--61~MeV above the threshold, which -- given the accuracy of the model -- does not exclude a weakly
 bound state. 
 This is mainly due to the fact that the hyperfine splitting between spin 1 and spin 0 $cb$ diquarks
 is only 10~MeV, while the difference of pertinent thresholds is 45~MeV. 
The fact that  $0^+$ $cb$ teraquark could be  bound was raised in Ref.~\cite{Weng:2021hje}.

\subsection{Sextet Masses}

In the case of  sextet tetraquarks, we have several spin states, since the soliton spin is $J=1$ and the 
$QQ$ diquark spin is  $S_{QQ}= 1$, and additionally  0 in the case of the $cb$ diquark. However, the pertinent spin splittings
are very small. Indeed, for the $bc$ diquarks spin splitting
is of the order of 10 MeV (see Fig.~\ref{fig:diqcb}) and the diquark-soliton spin splitting, depending on the diquark mass, is of
the order of 60, 20 and 15~MeV for $cc$, $bc$ and $bb$ tetraquarks.

Therefore, in the following we show only some representative plots for non-strange sextet tetraquarks. For 
tetraquarks with nonzero strangeness these curves have to be shifted upwards by the mass difference
between heavy baryons used as a reference -- see Eq.~(\ref{eq:tetra6}), and the pertinent thresholds have to be
replaced by the ones from Table~\ref{tab:thresholds}.

\begin{figure}[h]
\centering
\includegraphics[width=8.2cm]{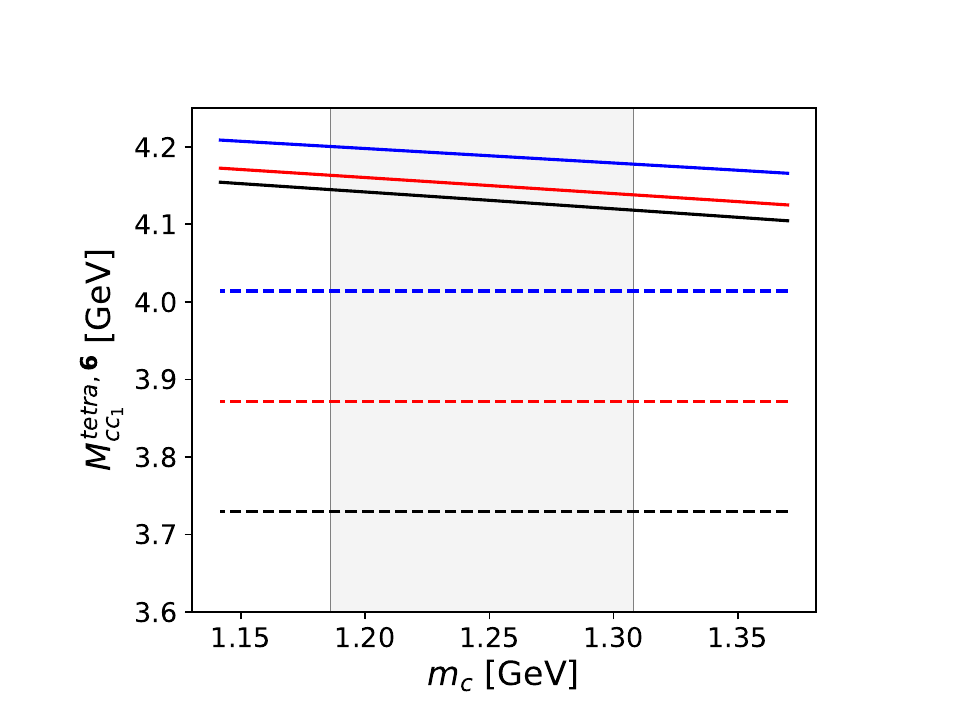} \\
\includegraphics[width=8.2cm]{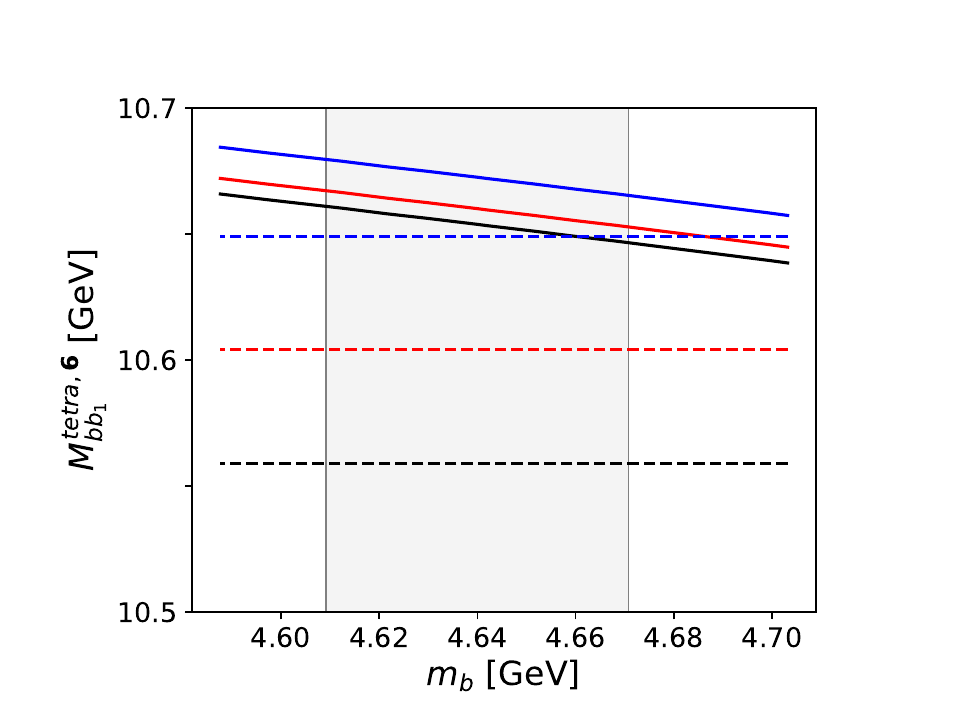}
\vspace{-0.2cm}%
\caption{Masses of $T_{cc}$ (upper panel) and $T_{bb}$ (lower panel) non-strange sextet tetraquarks of spin 0, 1
and 2 (from bottom upwards)
as functions of $m_Q$. Dashed lines show pertinent thresholds.
Vertical line indicates the value
of $m_Q$ corresponding to $\sigma$ for which  $2S$ mesons are best reproduced.  
Shaded area corresponds
to the limits of Eq.~(\ref{eq:qlimits}).
}
\label{fig:T6ccbb}%
\end{figure}

In Fig.~\ref{fig:T6ccbb} we plot non-strange $cc$ and $bb$ tetraquark masses. In this case tetraquarks have spin 0,1 or 2
and they are shown by different colors: blue, orange and green (from bottom to top), respectively. Pertinent thresholds
are marked by dashed lines. In both cases no bound states exist. These results are in agreement with our previous
estimates from Ref.~\cite{Praszalowicz:2022sqx}.

In Fig.~\ref{fig:T6ccbb} we plot the non-strange $cb$ tetraquark mass for diquark of spin 0, therefore the tetraquark spin is $s=1$. We see again
that two different mass estimates based on $c$ or $b$ baryons  in Eq.~(\ref{eq:tetra6}) differ by $15 \div 95$~MeV. In order
to illustrate the pattern of spin splittings we show in Table~\ref{tab:MT6cb} predictions for all spin combinations at the aggregate
mass $m_c+m_b=5.932$~GeV. We see that the spin splittings at this mass are of the order of $10$~MeV, whereas
the uncertainty due to the reference baryon (charm or bottom) is of the order of 50~MeV. All states are above the threshold.

\begin{figure}[h]
\centering
\includegraphics[width=7.7cm]{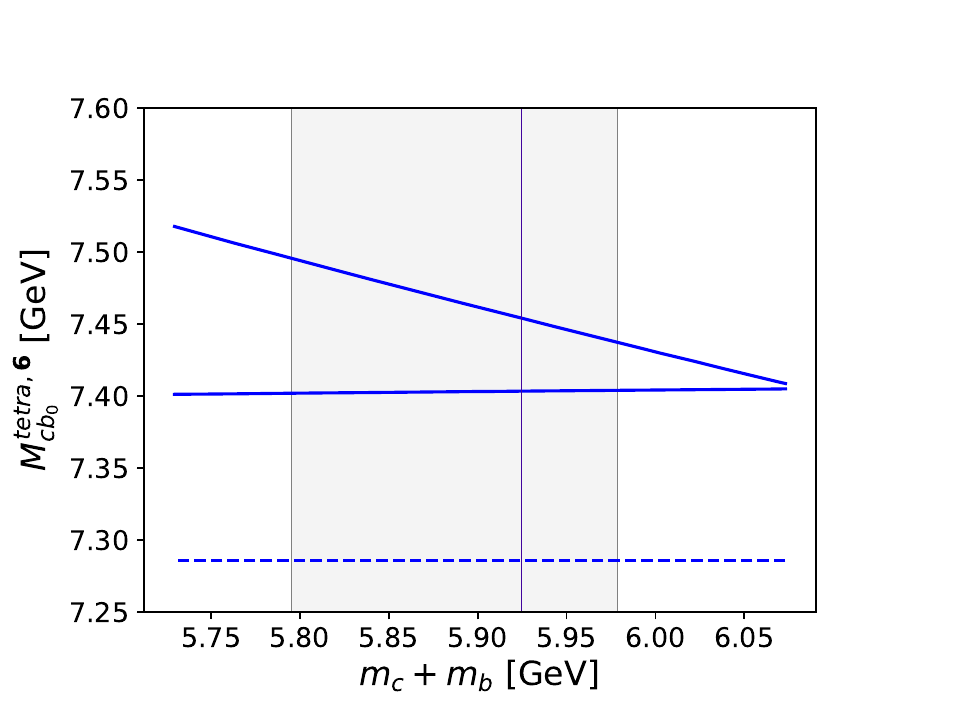} 
\vspace{-0.2cm}%
\caption{Mass of $T_{{c}{b}}$  non-strange sextet tetraquark of spin 0
as function of $m_c+m_b$. 
Upper line corresponds to the mass computed from the charm baryon spectrum, whereas the 
lower line to the bottom baryon (\ref{eq:tetra6}).
Dashed line shows the pertinent threshold.
Vertical line indicates the value
of $m_c+m_b$ corresponding to $\sigma$ for which  $2S$ $B_c$ meson is best reproduced.  
Shaded area corresponds
to the limits of Eq.~(\ref{eq:qlimits}).
}
\label{fig:T6cb}%
\end{figure}

\renewcommand{\arraystretch}{1.3} 
\begin{table}
\begin{tabular}[c]{|c|c|c|}
\hline
 $S_{\bar{c}\bar{b}}$ & 0 & 1 \\
 \hline
 $s=0 $ & $\cdots$ &$ 7.39 - 7.44$ \\
 \hline
 $s=1$ & $7.40 - 7.45$ & $7.40 - 7.45$ \\
 \hline
 $s=2$ & $\cdots$ & $7.42 -7.47$ \\
 \hline
\end{tabular}
\caption{Masses (in GeV) of non-strange sextet $cb$ tetraquarks of spin $s$ for $\bar{c}\bar{b}$ diquark of spin $S_{\bar{c}\bar{b}}$.}
\label{tab:MT6cb}
\end{table}
\renewcommand{\arraystretch}{1.0}

\section{Summary and Conclusions}
\label{sec:summary}

In this work, we have calculated the masses of heavy tetraquarks in a model in which the light sector is described 
by a chiral quark-soliton model, while the mass of the heavy diquark is calculated from the Schrödinger equation
with the Cornell potential including spin interactions. 
In this way, we extended our previous analysis \cite{Praszalowicz:2022sqx}
where the explicit spin interaction was ignored and only tetraquarks with identical heavy antiquarks were considered.
Since we have been interested in  $1S$ ground states only,
there was no need to include tensor and spin-orbit couplings. We have developed our own fitting procedure
to fix the parameters of the Cornell potential, including heavy quark masses, from the charmonium and bottomium
spectra. The resulting quark masses are in agreement with the quark masses extracted from the heavy baryon spectra
calculated in the framework of the $\chi$QSM, see Fig.~\ref{fig:mQvssig}. 
This proves the consistency of our approach.
Moreover, our parameters are in a reasonable agreement with the results from the global fits \cite{Mateu:2018zym}.

Furthermore, having all parameters fixed, we have calculated $B_c$ meson masses with no additional input. These predictions
agree very well with two cases known experimentally, see Eq.~(\ref{eq:Bcmasses}). This reassured us that  the parameters of the 
Cornell potential were correctly extracted from the $\bar{c}c$ and $\bar{b}b$ spectra, and that the interpolation to the $\bar{c}b$
system was correctly performed. We have also predicted masses of $B_c$ vector mesons (\ref{eq:Bcmasses1}).

In order to compute diquark masses we have rescaled appropriately the color factors entering the Cornell potential,
since the two antiquark color charges couple in this case to an SU(3) triplet  rather than to a singlet. The results are given
in Sect.\ref{ssec:numQQ}.

Heavy tetraquarks can be characterized according to the SU(3) classification of heavy baryons,
in which the heavy quark $Q$ has been replaced by
an anti-diquark. Mass formulas (\ref{eq:tetra3bar}) and (\ref{eq:tetra6})
include therefore heavy baryon masses and diquark masses, and the spin-spin interaction.
They are analogous to the phenomenological mass formulas of Ref.~\cite{Eichten:2017ffp}.

Our main conclusion is that only the $bb$ tetraquarks are bound, both non-strange and strange.
Unfortunately, the $cb$ system is not heavy enough to create a bound state. One of the
main motivations of the present paper was to check, whether $cb$ teraquarks exist.
There is still a possibility that strange $cb$ teraquark mght exist, since -- given the accuracy of the present approach -- our predictions
lie very close to the $\bar{D}^{*0}\bar{B}^0_s$ threshold.

In view of these findings it seems likely that the LHCb charm tetraquark is a kind of molecular configuration
\cite{Janc:2004qn,Li:2012ss}, which is beyond our present approach.

\section*{Acknowledgements}

This research has been Supported by the Polish National Science Centre Grant
2017/27/B/ST2/01314.


\end{document}